\documentstyle[aps,prl,twocolumn,epsfig]{revtex}

\begin{document}
\twocolumn[\hsize\textwidth\columnwidth\hsize\csname
@twocolumnfalse\endcsname

\title{Interdependence of Magnetism and Superconductivity in the Borocarbide
TmNi$_2$B$_2$C}

\author{K. N\o rgaard, M. R. Eskildsen\cite{dpmc} and N. H. Andersen}
\address{Condensed Matter Physics and Chemistry Department,
         Ris\o \ National Laboratory, DK-4000 Roskilde,
         Denmark}

\author{J. Jensen, P. Hedeg\aa rd and S. N. Klausen}
\address{\O rsted Laboratory,
         Niels Bohr Institute fAPG,
         Universitetsparken 5, DK-2100 Copenhagen, Denmark}

\author{P. C. Canfield}
\address{Ames Laboratory and Department of Physics and Astronomy,
         Iowa State University, Ames, Iowa 50011}

\date{December 14, 1999}
\maketitle

\begin{abstract}We have discovered a new antiferromagnetic phase in
TmNi$_2$B$_2$C by neutron diffraction. The ordering vector is
$\bbox{Q}_A = (0.48,0,0)$ and the phase appears above a critical
in-plane magnetic field of 0.9 T. The field was applied in order
to test the assumption that the zero-field magnetic structure at
$\bbox{Q}_F = (0.094,0.094,0)$ would change into a $c$-axis
ferromagnet if superconductivity were destroyed. We present
theoretical calculations which show that two effects are
important: A suppression of the ferromagnetic component of the
RKKY exchange interaction in the superconducting phase, and a
reduction of the superconducting condensation energy due to the
periodic modulation of the moments at the wave vector
$\bbox{Q}_A$.
\bigskip

\end{abstract}
 ]

The interplay between magnetism and superconductivity  is
inherently of great interest since the two phenomena represent
ordered states which are mutually exclusive in most systems.
Therefore, the borocarbide intermetallic quaternaries with
stoichiometry (RE)Ni$_2$B$_2$C have attracted great attention
since the publication of their discovery in 1994
\cite{nagara94,cava94}, as they exhibit coexistence of magnetism
and superconductivity if the rare earth (RE) is either Dy, Ho, Er
or Tm. The magnetic moments in these compounds are due to the
localized $4f$ electrons of the rare-earth ions. The $4f$ and the
itinerant electrons are coupled weakly by the exchange
interaction, resulting in the indirect
Ruderman-Kittel-Kasuya-Yoshida (RKKY) interaction between the
$4f$-moments. Thus the RKKY interaction, which is decisive for the
cooperative behavior of the magnetic electrons, depends on the
state of the metallic ones.

The borocarbides have a tetragonal crystal structure with space
group (I4/{\em mmm}) \cite{siegri94}, and TmNi$_2$B$_2$C has a
superconducting critical temperature $T_c^{} = 11$ K and a N\'eel
temperature $T_N^{} = 1.5$ K \cite{movsho94}. The crystalline
electric field aligns the thulium moments along the $c$ axis, and
the magnetic structure has a short fundamental ordering vector
$\bbox{Q}_F = (0.094,0.094,0)$ with several higher-order odd
harmonics \cite{lynn97,sternl97}. In the magnetic structures
detected in any of the other systems, the moments of the
rare-earth ions are confined to the basal plane, and they have
short wavelength antiferromagnetically ordered states. For
example, they are commensurate with a propagation vector $\bbox{Q}
= (0,0,1)$ for RE = Ho and Dy, or incommensurate with $\bbox{Q}
\approx (0.55,0,0)$ for RE = Gd, Tb, Ho and Er \cite{lynn97}. An
especially tight coupling between magnetism and superconductivity
has been clearly demonstrated in TmNi$_2$B$_2$C, where a magnetic
field applied along the $c$ axis induced concurrent changes of the
magnetic structure and the symmetry of the flux line lattice
\cite{eskild98}. Similar effects have not yet been observed in any
of the other borocarbides.

One question that immediately arises, and which we believe is
important for the general understanding of the interaction between
superconductivity and magnetism, is why the long-period magnetic
ordering found in TmNi$_2$B$_2$C is stable. Bandstructure
calculations on the normal state of non-magnetic LuNi$_2$B$_2$C
predict a maximum in the conduction-electron susceptibility
$\chi(\bbox{q})$ at $\bbox{q}\approx(0.6,0,0)$ supported by
Fermi-surface nesting \cite{rhee95,dugdal99}. The RKKY interaction
is proportional to the magnetic susceptibility of the electron
gas, and the position of its maximum, determining the magnetic
ordering vector, is expected to be nearly the same as in
$\chi(\bbox{q})$. This is in agreement with the experimental
findings for many of the magnetic borocarbides, but not for
TmNi$_2$B$_2$C.

The use of the band structure of the normal state ignores that the
magnetic interactions are mediated by a superconducting medium.
The BCS ground state is a singlet and apart from the interband
scattering the electronic susceptibility is therefore zero at
$\bbox{q}=\bbox{0}$ and zero temperature. It will recover its
normal-state value only when $q\gtrsim 10\,\xi^{-1}$, where $\xi$
is the coherence length of the superconductor, as first pointed
out by Anderson and Suhl \cite{anders59}. This raises the
possibility that a local maximum in the susceptibility is shifted
from zero in the normal phase to a non-zero value of $q$ in the
superconducting phase. In the case of the free electron gas, this
maximum is very shallow and occurs at a relatively large $q$. If,
however, interband effects (or umklapp processes in the free
electron model) are included the situation is changed. We shall
write the RKKY interaction in the superconducting phase as ${\cal
J}(\bbox{q}) = I[{\chi}_0^s(\bbox{q}) + {\chi}_u^{}(\bbox{q})]$.
${\chi}_0^s(\bbox{q})$ is the contribution from the intraband
scattering near the Fermi surface, which is sensitive to the
superconducting energy gap. We have determined this function
numerically in the zero temperature limit by using a linear
expansion in $q$ of the band electron energies near the Fermi
surface. A simple fit to the result is ${\chi}_0^s(q) =0.99\,q\,[q
+ 1.5 \xi^{-1}]^{-1}_{} {\chi}_0^n(0)$, where
$\xi=\pi\Delta/(\hbar v_F)$ is the coherence length. The
superconducting energy gap is negligible compared to the band
splittings, and the contribution ${\chi}_u(\bbox{q})$ from the
interband scattering is therefore not affected by the state of the
conduction electrons. At small $q$ we may assume: ${\chi}_u^{}(q)
= {\chi}_0^n(0)(\alpha- A q^2)$, where $A$ is a constant
normal-state quantity, which is expected to be some few times
$(2\pi/a)^{-2}$, where $a$ is the lattice parameter. The local
maximum in ${\cal J}(q)$, at $q=0$ in the normal phase, will in
the superconducting phase appear at $q_0 \approx [(4/3) A
\xi]^{-1/3}$. This is the same dependence on $\xi$ as found by
Anderson and Suhl, but the coefficient does not depend explicitly
on $k_F$ and is somewhat smaller. With $A(2\pi/a)^2$ lying between
1 and 10, $q_0$ assumes a value between 0.084 and 0.18 times
$2\pi/a$, which is consistent with the magnitude of the observed
ordering vector, $Q_F =0.13 (2\pi/a)$.

The existence of a local maximum in the RKKY interaction is not
sufficient for stabilizing a magnetic structure with that
wavelength. The total free energy of the entire system, including
the condensation energy of the superconducting electrons, has to
be minimal. The presence of a magnetic modulation at $\bbox{Q}$
may influence and weaken the superconductivity, for instance
through a reduction of the density of states at the Fermi surface
due to the superzone energy gaps created by the periodic
modulation of the moments. This effect depends strongly on
$\bbox{Q}$ and may be an important factor in the competition
between alternative magnetic structures.

In this letter we report the results of neutron diffraction
studies on TmNi$_2$B$_2$C augmented by theoretical calculations of
the competition between the magnetic and superconducting states.
With the objective of suppressing superconductivity while the
$c$-axis components of the magnetic moments are still ordered, a
magnetic field was applied in the basal plane. This allows a study
of the magnetic state as it would be in the absence of
superconductivity. The experimental results are described in
detail below. Briefly we found that the system does not form a
ferromagnetic state when superconductivity is suppressed, as one
might expect from the arguments presented above, but instead it
relaxes into an incommensurate antiferromagnetic state with
$\bbox{Q} = \bbox{Q}_A = (0.482,0,0)$, resembling the magnetic
ordering most commonly found in the borocarbides.

The experiments were performed on a $2\times 2\times 0.2$ mm$^3$
single crystal of TmNi$_2$B$_2$C grown by a high-temperature flux
method and isotopically enriched with $^{11}$B to enhance the
neutron transmission \cite{cho95a,cho95b}. The sample was mounted
on a copper rod in a dilution refrigerator insert for the
low-temperature measurements at $T < 1.7$ K, and on a standard
sample stick for measurements above 1.7 K. In both cases the
sample was oriented with the $a$- and $b$-axes in the scattering
plane and placed in a $1.8$ T horizontal-field magnet aligned
along the $a$ axis. Measurements were performed with applied
fields up to the maximum of $1.8$ T, and at temperatures between
20 mK and 6 K. The neutron diffraction experiments were performed
at the TAS7 triple-axis spectrometer on the cold neutron beam line
at the DR3 research reactor at Ris\o\ National Laboratory. The
measurements were carried out with $12.75$ meV neutrons,
pyrolythic graphite (004) monochromator and (002) analyzer
crystals, and in an open geometry without collimation. The
effective beam divergence before and after the sample was 60 and
120 arc minutes, respectively.

\centerline{\epsfig{file=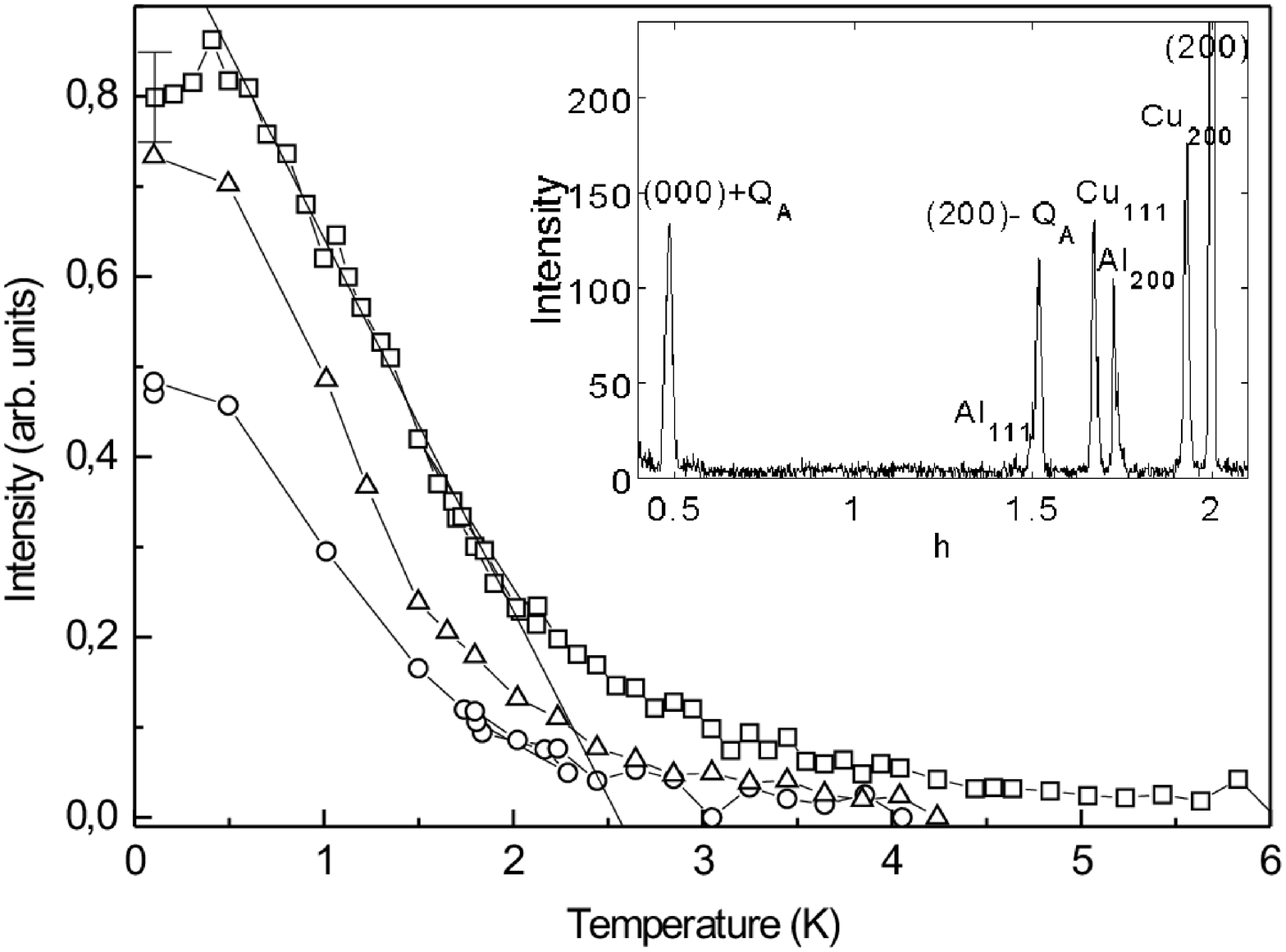, width = 1.13\linewidth,
clip=}}\vspace{-5pt} {\footnotesize\noindent{\bf Fig.~1}
Normalized integrated intensities versus temperature of the
field-induced magnetic peaks at $1.2$ T (circles), $1.4$ T
(triangles) and $1.8$ T (squares). The linear fit to the $1.8$ T
data shows the determination of $T_N^{}(B)$. Inset: Scan along the
$[h\,0\,0]$ direction at $T = 100$ mK and $1.8$ T, showing the
field-induced satellite peaks at $\bbox{Q}_A = (0.482,0,0)$ around
the $(0\,0\,0)$ and $(2\,0\,0)$ nuclear reflections. The peak
intensity of the $(2\,0\,0)$ reflection is
800.\par\baselineskip10pt}
\bigskip

In the inset to Fig.~1 is shown the result of a scan along $[1 \,
0 \, 0]$ at $T = 100$ mK and $B = 1.8$ T. In addition to the
nuclear $(2 \, 0 \, 0)$ Bragg reflection, the figure shows
satellite peaks at $(0\,0\,0) + \bbox{Q}_A$ and $(2\,0\,0) -
\bbox{Q}_A$ with $\bbox{Q}_A = (0.482,0,0)$. These satellite peaks
are not observed in zero applied field neither below nor above
$T_N^{}$, and they do not appear until the field is larger than
0.9 T at 100 mK. Additional magnetic satellites were detected at
$(1\,1\,0) \pm \bbox{Q}_A$ and $(0\,2\,0) \pm \bbox{Q}_A$. No
higher order harmonics of the field-induced satellites were
observed, and their intensities are consistent with the magnitude
and direction of the magnetic moments remaining equal to
$3.8~\mu_B^{}$ and being essentially parallel to the $c$ axis. At
all temperatures and fields, where the $\bbox{Q}_A$ peaks were
observed, they stayed resolution limited. Likewise no field or
temperature dependence of $\bbox{Q}_A$ was found. Quite
remarkably, the field-induced peaks are only observed for the wave
vector $\bbox{Q}_A \parallel \bbox{B}$ and not for $\bbox{Q}_A
\perp \bbox{B}$, hence the lowering of the in-plane four-fold
symmetry of the system produced by the field has a direct
consequence on the direction of $\bbox{Q}_A$.

\bigskip \centerline{\epsfig{file=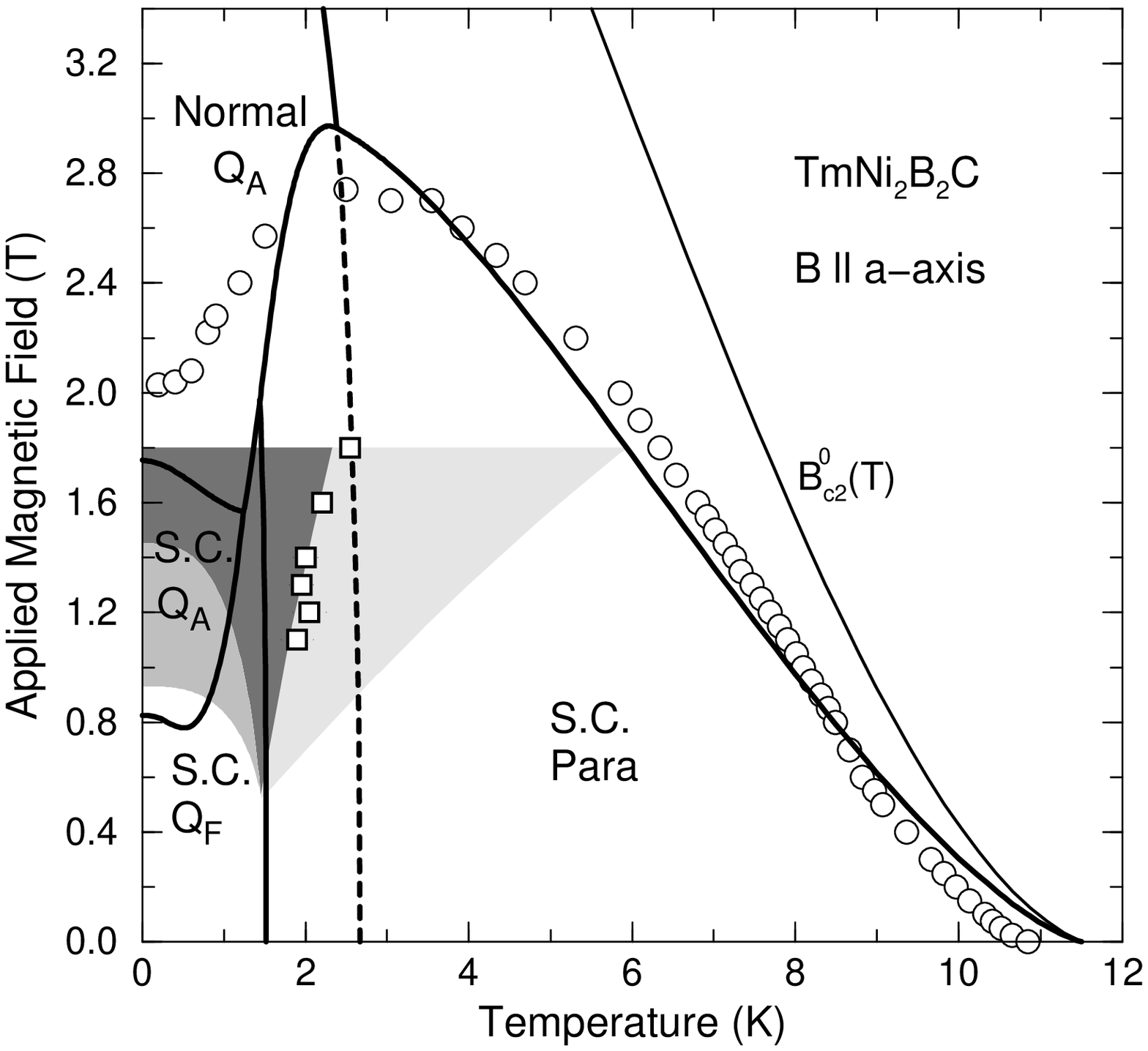, width =
0.95\linewidth, clip=}} {\footnotesize\noindent{\bf Fig.~2}
Experimental and theoretical phase diagram of TmNi$_2$B$_2$C in a
magnetic field along $[1\, 0\, 0]$. The medium-gray area denotes
the region where both the $\bbox{Q}_A$ and the $\bbox{Q}_F$
reflections are present. In the dark-gray area only the
$\bbox{Q}_A$ reflections were observed, up to the maximum applied
field of 1.8 T. The light-gray area denotes the region where the
long tail of low-intensity magnetic scattering at $\bbox{Q}_A$ is
still observed. The squares denote the measured phase boundary
between the $\bbox{Q}_A$ phase and the paramagnetic one,
$T_N^{}(B)$, determined by the procedure described in the text.
The open circles denote the upper critical field determined by
transport measurements \cite{naugle99}. The solid lines are the
theoretical phase boundaries. The dashed line is the calculated
N\'eel temperature of the $\bbox{Q}_A$ phase had the metal stayed
in the normal state. The thin line labeled $B_{c2}^0(T)$ is the
estimated upper critical field if the magnetic subsystem is
neglected.\par \baselineskip10pt} \bigskip

Concurrently with the appearance of the field-induced peaks the
intensity of the zero-field magnetic reflections with scattering
vector $\bbox{Q}_F$ decreases and finally vanishes at $1.4$ T and
100 mK. Between 0.9 and $1.4$ T the magnetic structures at
$\bbox{Q}_F$ and $\bbox{Q}_A$ coexist. The length of $\bbox{Q}_F$
does not change for applied magnetic fields up to 0.9 T. Above 0.9
T a small reduction of $|\bbox{Q}_F|$ is observed, simultaneously
with the appearance of the field-induced magnetic reflection at
$\bbox{Q}_A$. The reduction is at the most $3\%$, just before the
peaks vanish at $1.4$ T.

In the main body of Fig.~1 we show the temperature dependence of
the integrated intensity of the field-induced magnetic reflections
for three different values of the applied fields. This shows that
the intensity of the peaks increases with increasing field at a
constant temperature. The results at the different values of the
field show qualitatively the same temperature dependence, a rapid
linear decrease when the temperature is above $\sim0.5$ K and a
cross-over into a long tail at about 2 K. At the maximum field of
1.8 T the tail extends up to a temperature of four times the
zero-field value of $T_N^{}$. For each value of the field
$T_N^{}(B)$ is defined as the extrapolation to zero of the linear
part of the integrated intensity, as shown in the figure for the
case of 1.8 T. The experimental data are summarized in the phase
diagram in Fig.\ 2.

The theoretical phase diagram is calculated using the following
parameters: 1) The crystal-field Hamiltonian of the Tm ions
determined from experiments \cite{gasser96,jens91}. 2) A
phenomenological RKKY interaction ${\cal J}(\bbox{q})$ with two
parameters ${\cal J}(\bbox{Q}_F)$ and ${\cal J}(\bbox{Q}_A)$. The
normal-state value of ${\cal J}(\bbox{0})\approx{\cal
J}(\bbox{Q}_F)$. 3) Abrikosov's formula \cite{Tink} for the
condensation energy of the superconducting state
\begin{equation}
F_s-F_n =
-\{B_{c2}^0(T)-B_i^{}\}^2[1.16\cdot8\pi(2\kappa_{}^2-1)]^{-1},
\end{equation}
where $B_i$ is the internal magnetic field, corrected for the
uniform magnetization and demagnetization effects. In the present
system $B_i$ is about 10\% larger than the applied field.
$B_{c2}^0(T)$ is the upper critical field if the coupling to the
magnetic electrons is neglected. Its dependence on $T/T_c$ is
assumed to be the same as observed in the non-magnetic Lu
borocarbide \cite{shulga98}. $B_{c2}^0(T=0)$ has been used as a
fitting parameter, and its final value of 6.5 T (corresponding to
$\xi =71$ \AA) is close to that obtained when scaling the values
of $B_{c2}^0(0)$ in the non-magnetic Lu and Y borocarbides with
$T_c$. The other fitting parameter in Eq.\ (1) is the
(renormalized) value of $\kappa$ which is found to be 6.3, close
to that determined experimentally by Cho {\it et al.}
\cite{cho95a}. 4) The coupling between the magnetic system and the
superconducting electrons is described by two parameters: a) The
Anderson--Suhl reduction of ${\cal J}(\bbox{0})$ in the
superconducting phase at zero temperature, which is found to be
close to the value of ${\cal J}(\bbox{0})$ itself. The thermally
excited quasiparticles implies that the reduction is smaller at
finite temperatures. This effect is included in the calculations.
b) The suppression of the superconductivity, which is assumed to
be due to the superzone energy gaps near the Fermi surface
produced by the magnetic ordering at $\bbox{Q}_A$. The density of
states at the Fermi surface is reduced proportionally to the sizes
of the energy gaps, which by themselves are proportional to the
amplitude of the magnetic modulation \cite{elliott63}. The
reduction of the density of states causes a decrease of
$B_{c2}^0(T)$ in Eq.\ (1), which is estimated to be at most 40\%.

The model calculations account reasonably well for the
experimental phase boundaries in Fig.\ 2, and equally well for the
phase diagram when the field is applied in the $c$ direction, as
measured by Eskildsen {\it et al.}\ \cite{eskild98}. The large
anisotropy between the experimental upper critical field along the
$a$ and along the $c$ axis is not an internal property of the
superconducting electron system, but is due to the large
difference between the magnetic $a$- and $c$-axis susceptibility
of the Tm ions. Although the transformation of the $\bbox{Q}_F$
state into the $c$-axis ferromagnet did not occur, it is clear
that the Anderson--Suhl mechanism plays an important role for the
behavior of TmNi$_2$B$_2$C, see also the discussion by Kuli\'c
{\it et al.} \cite{kulic97}. The loss in the superconducting
condensation energy at the field-induced transition to the normal
state is compensated for by the gain in the RKKY exchange energy
deriving from the uniform magnetization because of the sudden
increase of ${\cal J}(\bbox{0})$. This mechanism explains the
large reduction of the upper critical field shown in Fig.\ 2, and
the even larger reduction in the $c$-axis phase diagram. Most
significantly, it explains why the $\bbox{Q}_F$ state is stable up
to a field of 1 T applied along the $c$ axis, i.e.\ as long as the
system stays superconducting. In case of a normal behavior of
${\cal J}(\bbox{q})$, the $c$-axis field would destroy this phase
almost immediately.

Against intuition, the short wavelength magnetic modulation in the
$\bbox{Q}_A$ phase affects the superconductivity state much more
strongly than the long wavelength $\bbox{Q}_F$-modulation, but it
should be noticed that the wavelengths of the two structures are
both much shorter than $\xi$. The model predicts that in a normal
magnetic system the $\bbox{Q}_A$ phase would be the stable one at
low temperatures and fields. However, the superzone energy gaps
produced by the $\bbox{Q}_A$-modulation disturb the nesting
features of the Fermi surface close to this wave vector, causing a
strong suppression of the superconductivity in the Tm system. This
effect is also found in the measurements on e.g.\ Ho based
borocarbides, where superconductivity is suppressed when the
magnetic system enters the $\bbox{Q}=(0.55,0,0)$ phase and is
regained when the system leaves this phase at a yet lower
temperature \cite{schmidt97}. Er borocarbide is the only one where
the $(0.55,0,0)$-phase does not seem to strain the superconducting
properties severely, although the upper critical field
\cite{cho95b} shows similarities with the observations in the Tm
system. The reason that the $\bbox{Q}_F$ phase does not appear in,
for instance, the Er system may simply be that the RKKY
interaction, being proportional to $[(g-1)J]^2$, is stronger in Er
than in Tm making the energy difference between the two different
magnetic phases too large in comparison with the superconducting
condensation energy.

Improvement of the present theory is planned in order to account
for the inhomogeneities in the superconducting order parameter due
to the flux lines. It is difficult to understand how the
low-intensity long tail of the $\bbox{Q}_A$ reflections may extend
up to a temperature which is twice the estimated transition
temperature of the $\bbox{Q}_A$-phase in a normal system (the
dashed line in Fig.\ 2). The phase may survive within the normal
core of the flux lines, but not very far above the normal-phase
transition temperature. Further experiments are planned at higher
fields in order to test the theoretical predictions, e.g.\ that
$T_N^{}(B)$ coincides with the maximum in the in-plane upper
critical field, to investigate in more detail the magnetic
scattering in the long-tail regime, and to search for the
$\bbox{Q}_A$-phase in the $c$-axis phase diagram.

In this letter we have proposed that the long wavelength magnetic
structure of TmNi$_2$B$_2$C below 1.5 K owes its very existence to
the fact that the material is superconducting. This was shown
experimentally by applying an in-plane magnetic field, which in
this Ising-like system primarily affects the superconducting
state. We have found that the magnetic system enters a new phase
at a critical field of 0.9 T where the order is at short
wavelength, $\bbox{Q}_A = (0.48,0,0)$. We have presented
theoretical calculations showing that the interplay between
superconductivity and magnetism in the borocarbides is governed by
two mechanisms: 1) a suppression of the ferromagnetic component of
the RKKY exchange interaction in the superconducting phase, and 2)
a reduction of the superconducting condensation energy from the
periodic modulation of the moments at the wave vector
$\bbox{Q}_A$.

We thank D. G. Naugle and K. D. D. Rathnayaka for sharing their
data prior to publication. This work is supported by the Danish
Technical Research Council and the Danish Energy Agency. P.C.C. is
supported by the Director of Energy Research, Office of Basic
Energy Science under contract W-7405-Eng.-82.

\end{document}